\documentclass[twocolumn,aps,secnumarabic%
,showpacs,showkeys,preprintnumbers,%
amsmath,amssymb]{revtex4}

\usepackage{amsmath,amssymb}
\usepackage{float,epsfig}
\usepackage{graphicx}
\usepackage{dcolumn}
\usepackage{bm}

\begin{document}

\title{Structural Symmetry of Two-dimensional Metallic Arrays: Implications
for Surface Plasmon Excitations}

\author{H. Iu}
\affiliation{Department of Physics,
The Chinese University of Hong Kong,
Shatin, New Territories, Hong Kong.}

\author{H. C. Ong}
\affiliation{Department of Physics,
The Chinese University of Hong Kong,
Shatin, New Territories, Hong Kong.}

\author{Jones T. K. Wan}\email{jwan@phy.cuhk.edu.hk}
\affiliation{Department of Physics,
The Chinese University of Hong Kong,
Shatin, New Territories, Hong Kong.}

\date{\today}

\begin{abstract}
In recent years, there has been intensive investigation of surface plasmon polaritons (SPPs)
in the science and engineering fields.  Understanding the physics of
surface plasmon excitation is essential to the manipulation of SPPs, and most existing
studies focus on ($-1,0$)-type SPP excitation.  In this article, we
report our recent investigation of the ($0,\pm 1$)-type SPP excitation of a gold
two-dimensional nano-cavity array using finite-difference time-domain
methodology.  Our particular focus is on the symmetry properties of
($0,\pm 1$)-SPPs excited by different polarizations of light.  It is found
that polarization has strong implications for the field distribution of the
corresponding SPPs.  As a result, the control of polarization may provide
important insights into the manipulation of SPPs.
\end{abstract}

\maketitle

\section{Introduction}
Recent advances in nanotechnology have ignited the study of surface
plasmon polaritons (SPPs), which have become a popular topic
\cite{garcia_de_abajo_rmp_2007, barnes_nature_2003, raether_1988} due to their
wide range of potential applications in surface-enhanced Raman scattering (SERS)
\cite{kneipp_2006}, surface enhanced second harmonic generation,
nano-photonics \cite{lai_nat_photon_2007}, thermal-photovoltaic devices
\cite{wilde_nature_2006, billaudeau_apl_2008,
greffet_nature_2002,wan.opt.commun.2009}, and data storage and imaging
\cite{schmidt_prb_2008}

However, finding a way to excite and control SPPs in an advantageous manner is still the
main concern of scientists.  The technology available to control SPPs with
precision and flexibility remains underdeveloped, and different schemes
have been proposed in this regard.  For example, since the
discovery of extraordinary transmission \cite{ebbesen_nature_1998}, major
interest has centered on investigating cylindrical hole arrays in which
subwavelength hollow cylinders are periodically formed on a flat metal film
using lithographic methods \cite{barnes_prl_2004,popov_prb_2000}.
The work of \citet{kelf_prl_2005}, however, has demonstrated  showed that the shape of
an individual cavity in metallic grating plays a dominant role in controlling
the excitation of SPPs.  In addition,  \citet{molen_2005}  showed that there are
shape and localized resonances in two-dimensional (2D) periodic subwavelength
metallic cavity arrays.  As a result, different shapes can lead to different
resonances because the holes act like plasmonic cavities to confine the
electric field and thereby give rise to strong localized resonances
\cite{garcia_de_abajo_rmp_2007,moreno_joapao_2006,baida_prb_2003}.
Understanding the geometry effects of cavities can lead to a number of applications
that require the precise spatial and frequency control of an enhanced electric field,
such as SERS and thermal radiation.
However, scientists have only a limited understanding of the surface shape
resonances, and the correlation between SPP excitation and cavity geometry
has not been widely investigated until recently
\citep{kelf_prl_2005,sepulveda_2008,Sauvan_2008,chuang.opt.express.2008,li_apl_2008,iu_2008,li.aspect.apl.2009,chen.opt.express.2009}.
In a recent work, we reported the fabrication of 2D nano-cavity arrays on
a gold surface  using interference lithography (IL) \cite{li_apl_2008}.  In this
article, we focus on a theoretical study of 2D nano-cavity arrays by using
finite-difference time-domain (FDTD) simulation \cite{taflove_2005}
methodology.  Our particular focus is on the symmetric properties and
polarization dependence of excited SPPs

\section{Simulation details}
\subsection{Basic simulation cell setup}
\begin{figure}[htbp]
\centerline{\epsfig{file=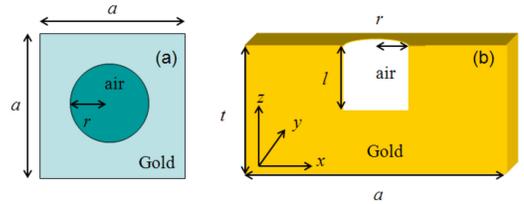,width=0.8\linewidth}}
\caption{(Color online) (a) Unit cell and (b) top view of the cylindrical cavity.}
\label{structure}
\end{figure}
We consider a model system that contains a gold slab drilled with a 2D
array of cylindrical cavities.  It is convenient to define the
unit cell shown in Fig. \ref{structure} with suitable boundary
conditions.  Periodic boundary conditions (PBCs) are applied in both the
$\pm x$- and $\pm y$-directions to produce a periodic structure.  To account
for non-periodicity in the $z$-direction, the perfectly matched layer (PML)
\cite{Berenge_1994} boundary condition is adopted in the $\pm
z$-directions. In addition, certain parameters are fixed throughout the
study. For example, we only examine optical wavelengths that range from $400$~nm to $900$~nm.
Moreover, the thickness of the gold layer is set at
$1$ $\mu$m to ensure that the structure can be considered to be optically thick and such bulk
photonic effects as Fabry-P\'{e}rot resonance and guiding mode resonance
\cite{fan_prb_2002} are substantially reduced and negligible.  Finally,
the spatial and time resolution are $\Delta x = 25$ nm and $\Delta t = \Delta
x/c$, respectively, and the two lattice constants are fixed at $a=575$~nm.

\subsection{Dielectric function of gold}\label{model_metal}
\begin{figure}[htbp]
\centerline{\epsfig{file=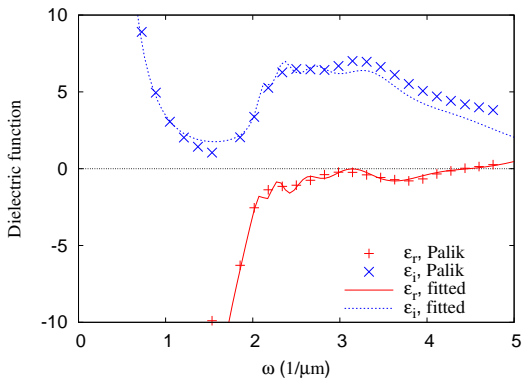,width=0.8\linewidth}}
\caption{(Color online) The real ($\varepsilon_{\rm r}$) and imaginary ($\varepsilon_{\rm
i}$) parts of the dielectric function fitted with the multiple Drude-Lorentz
model.}
\label{fit}
\end{figure}

Gold processes complex electron inter-band transitions at infrared
frequencies. As a result, simple dielectric functions such as the Drude model, which
considers only the electric response of conduction electrons, may not be
adequate for modeling the dielectric response of gold.  In this work, the
multiple Drude-Lorentz model proposed by \citet{vial_prb_2005} is adopted:
\begin{equation}
\tilde{\varepsilon}(\omega)=\varepsilon_{\infty}-\sum_{j}\frac{\triangle\varepsilon_{j}\omega_{j}^{2}}{\omega_{j}^{2}-\omega^{2}-i\omega\gamma_{j}}.
\label{Drude-Lorentz}
\end{equation}
Here, index $j$ represents the contribution to the dielectric function made by the
electrons of different bands; $\omega_{j}$ and $\gamma_{j}$ are the
frequency and damping parameters; $\varepsilon_\infty$ is the high-frequency
response, which is originated by the screening of the core electrons; and, finally,
$\Delta\varepsilon_j=\epsilon_\infty-\varepsilon_j$, where $\varepsilon_j$ is
the static dielectric response contributed by the different electron bands.
Equation~(\ref{Drude-Lorentz}) is fitted against the measured dielectric function
of gold \citep{palik_1985}, and the fitted parameters are tabulated in Table
\ref{table2}.  Figure \ref{fit} shows both the measured and fitted dielectric
functions.  As can been seen, Eq. (\ref{Drude-Lorentz}) agrees well with the
measured  $\varepsilon$ throughout our frequency range of interest.
\begin{table}[htbp]
\begin{center}
\begin{tabular}{|c|c|c|c|c|}
\hline $\varepsilon_{\infty}$ & $\triangle\varepsilon_{i}$ & $\omega_{i}$ $(1/\mu m)$ & $\gamma_{i}$ $(1/\mu m)$\\
\hline 5.339 & 6.2634$\times 10^{41}$ & 0.0486 & 1$\times 10^{-20}$\\
  & 0.1906 & 2.1201 & 0.1772\\
  & 0.9835 & 4.4214 & 1.9662\\
  & 1.5974 & 3.353 & 1.1844\\
  & 0.6653 & 2.7116 & 0.578\\
  & 0.4508 & 2.3525 & 0.3012\\
\hline
\end{tabular}
\end{center}
\caption{(Color online) Fitted parameters for Eq.~\ref{Drude-Lorentz}.} \label{table2}
\end{table}

\subsection{Dispersion relation calculation}
To calculate the dispersion relation, several Gaussian sources are placed
within the simulation cell.  These Gaussian sources have a frequency width that
covers the entire optical frequency range. They are randomly located so
that all of the resonant modes of the system can be excited.  After turning off
the sources, some fields are left in the system to allow their
magnitudes to be recorded as a finite-length signal.  This signal is then
expressed as the sum of a finite number of sinusoids with exponentially decaying
factors in a given bandwidth.  The frequencies, decay constants, amplitudes,
and phases of these sinusoids are determined from the coefficients of the sum.

Furthermore, line sources pointing in the $y$-direction are used to allow an investigation of
the effects of different polarizations on the dispersion relations.
These sources can be classified as $p$-polarized or $s$-polarized,
depending on whether it is the magnetic field ($H_{y}$) or electric field
($E_{y}$) that pointis in the $y$-direction, respectively.
\begin{figure}[htpb]
\centerline{\epsfig{file=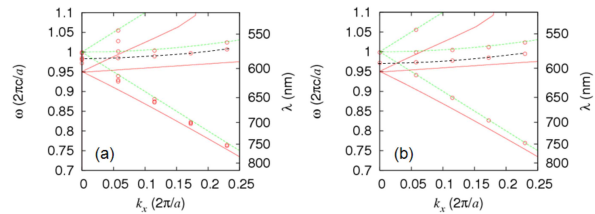,width=0.9\linewidth}}
\caption{(Color online) Dispersion relations of an $r=75$ nm cylindrical cavity
array under (a) $p$-polarized and (b) $s$-polarized lights. The green dotted
lines are Wood's anomalies [Eq. (\ref{wood})], the red lines are the SPP dispersions of
a gold slab [Eq. (\ref{spp.dispersion.relation})], and the black dotted lines are the
calculated ($0,\pm 1$) modes.}
\label{cylindrical:band:r75}
\end{figure}
Some of the resonant modes shown in Fig.~\ref{cylindrical:band:r75} correspond to
Wood's anomalies \citep{barnes_prl_2004}.  Wood's anomalies occur when one
diffracted order is parallel to the structure surface, which results in energy
redistribution to the other diffracted orders.  In other words, the incident light
is scattered in parallel to the structure surface.  This causes a sudden
change in reflectance when the frequency ($\omega$) and in-plane wave
vector ($k_{x}$) of the incident light satisfies
\begin{equation}\label{wood}
\omega = c | \vec{k} | 
=  c | \vec{k}_{x}+n\vec{G}_{x}+m\vec{G}_{y}|,
\end{equation}
where $\vec{k}$ is the incident wave-vector, $\vec{G}_{x}$ and $\vec{G}_{y}$
are the reciprocal lattice vectors, and $n$ and $m$ are integers.  Additionally,
we can trace the SPPs by plotting their dispersion relation driven by
the following equation
\begin{equation}\label{spp.dispersion.relation}
\omega = c
\sqrt{\frac{\varepsilon_{\rm Au}+\varepsilon_{\rm air}}{\varepsilon_{\rm Au}\varepsilon_{\rm air}}}|
\vec{k}_{x}+n\vec{G}_{x}+m\vec{G}_{y}|.
\end{equation}
It can be seen that Eq. (\ref{spp.dispersion.relation}) agrees well with the
remaining excited modes of the $p$-excited resonances for $(n,m)$ = $(-1,0)$,
$(1,0)$, and $(0,\pm 1)$, whereas only the $(0,\pm 1)$-type resonances are present for
$s$-polarization.  This is due to the fact that the array has $C_{4v}$ point
group symmetry. The eigenstates with $\vec{k}$ parallel to the
$\Gamma$-$X$ direction are irreducible representations of point group $C_{1h}$
\citep{sakoda_2001}, and each eigenstate is either symmetric ($A$ mode) or
antisymmetric ($B$ mode) under a mirror transformation along the $x$-$z$ plane
and can be excited only by incident waves that have the same symmetric property.
These are, as is discussed in Appendix \ref{symmetric.operation}, $p$- and
$s$-polarized lights, respectively.  Furthermore, it is proven that only
a symmetric eigenstate is present for ($n,m$) = ($-1,0$) and ($1,0$), but that, on the
contrary, both symmetric and antisymmetric eigenstates exist for ($n,m$) =
($0,\pm 1$).  Therefore, the black dotted lines shown in
Figs. \ref{cylindrical:band:r75}(a) and (b) represent the
eigenstates of the $A$ and $B$ symmetries, respectively.

\section{Results and discussion}\label{spp.excitation}
\subsection{Field pattern calculation}\label{field}
We are now in a position to present the field density distribution of the various
excited SPP modes.  Guided by the dispersion relations, we have located the
wave-vectors and frequencies of these excited modes at $k_{x}=0.1725$,
$\omega=0.9968$ for the $p$-excited resonance from
Fig. \ref{cylindrical:band:r75}(a) and $k_{x}=0.1725$, $\omega=0.9849$ for the
$s$-excited resonance from Fig.  \ref{cylindrical:band:r75}(b), respectively,
and have calculated their corresponding near field intensities.  To calculate
field intensity, a plane-wave Gaussian source is put above the array at
$z=1$~$\mu$m to mimic an incident beam.  The source has a narrow frequency
width, such that it simulates a single-frequency incident plane wave.  In
addition, a spatially dependent amplitude function, exp($2\pi i k_x x$), is
added to the source, where $k_{x}$ is the in-plane wave-vector, such that the
angle of incidence is given by $\sin^{-1}(k_{x}c/\omega)$.

This simulation provides us with useful information on the way in which the incident light
interacts with the array and the physical properties of the excited
resonances.  Time-domain simulations provide information about both the transient
and steady states.
We focus on the steady states by calculating the
spectral field density of an excited mode at specified $k_{x}$, $\omega$ values:
\begin{equation}
|\vec{E}(\vec{r},\omega)|^{2}=\int
(\vec{E}^{*}(\vec{r},t)\cdot\vec{E}(\vec{r},t)) e^{i\omega t}dt.
\end{equation}
Consider the time-average of energy density, as follow
\begin{eqnarray*}
& & \int(\vec{E}^{*}(\vec{r},t)\cdot\vec{E}(\vec{r},t)) dt\\
&=& \int(\int\vec{E}^{*}(\vec{r},\omega)e^{-i \omega t} d\omega\cdot\int\vec{E}(\vec{r},\omega ')e^{i \omega ' t} d\omega ')dt\\
&=& \int\int(\vec{E}^{*}(\vec{r},\omega)\cdot\vec{E}(\vec{r},\omega '))(\int e^{i(\omega '-\omega)t}dt)d\omega d\omega '\\
&=& \int\int(\vec{E}^{*}(\vec{r},\omega)\cdot\vec{E}(\vec{r},\omega '))2 \pi \delta(\omega '-\omega)d\omega d\omega '\\
&=& 2 \pi \int(\vec{E}^{*}(\vec{r},\omega)\cdot\vec{E}(\vec{r},\omega)) d\omega.
\end{eqnarray*}
Spectral field density is a single $\omega$ component of
$\vec{E}^{*}(\vec{r},\omega)\cdot\vec{E}(\vec{r},\omega)$.  Theoretically, we
need to sum up all of the $\omega$ components.  However, as the Gaussian source
used has only a narrow frequency width, $\Delta\omega$, the component of
the specified central frequency, $\omega_{\rm cen}$ dominates the integral.  By
assuming the other $\omega_{\rm others}$ components are weak, that is,
\begin{equation}
|\vec{E}(\vec{r},\omega_{\rm cen})|^{2}>>|\vec{E}(\vec{r},\omega_{\rm others})|^{2},
\end{equation}
we have
\begin{equation}
|\vec{E}(\vec{r},\omega_{\rm cen})|^{2}=\int\vec{E}^{*}(\vec{r},t)\cdot\vec{E}(\vec{r},t)dt.
\end{equation}
In addition, $ |\vec{E}(\vec{r},\omega_{\rm cen})|^{2}$ also provides
information about the harmonic eigenstate with a specified $k_{x}$ and
$\omega_{\rm cen}$.

\subsection{Symmetry properties of the excited SPPs}
\begin{figure}[htpb]
\centerline{\epsfig{file=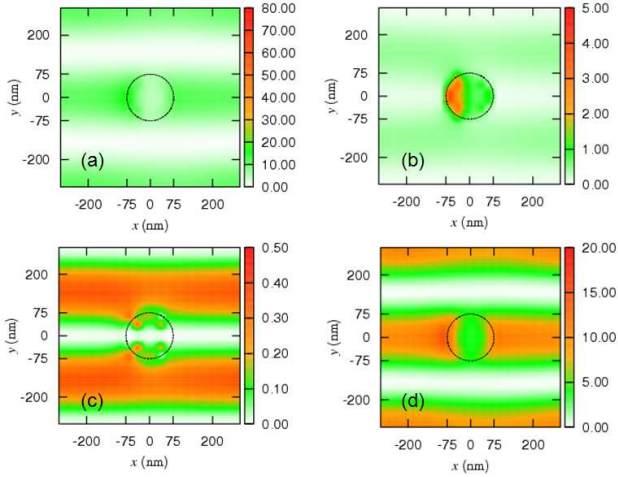,width=0.95\linewidth}}
\caption{(Color online) Electric field density on $z = 0$ of the ($0,\pm 1$)
SPP mode of an $r=75$ nm cavity array excited by a $p$-polarized light: (a)
$|E(\vec{r},\omega)|^2$, (b) $|E_x(\vec{r},\omega)|^2$, (c)
$|E_y(\vec{r},\omega)|^2$, and (d) $|E_z(\vec{r},\omega)|^2$.  The in-plane
wavevector and frequency of the excited SPP are $0.1275$ and $0.9968$,
respectively.  The fields are not drawn to scale for better visibility.  Note
the symmetry along $y=0$ and $y=\pm a/2$.}
\label{cylindrical:e:p:r75}
\end{figure}
\begin{figure}[htbp]
\centerline{\epsfig{file=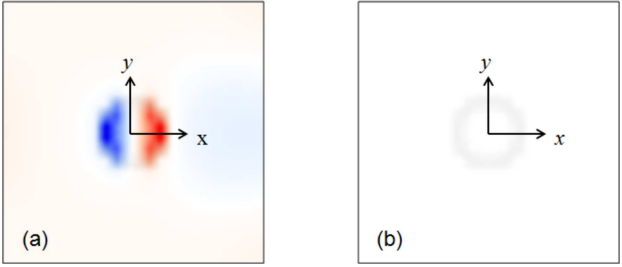,width=0.95\linewidth}}
\caption{(Color online) Snapshots of the electric field distributions,
$\vec{E}(\vec{r},t)$, for the $p$-excited SPP mode of
Fig. \ref{cylindrical:e:p:r75}, shown in (a) and (b), which are $E_x(\vec{r},t)$ and
$E_y(\vec{r},t)$, respectively ($E_y$ is weak and can barely be observed).  Note
the symmetry along $y=0$.  Blue and red represent positive and negative
values, respectively.}
\label{cylindrical:exy:field:p:r75}
\end{figure}
\begin{figure}[htpb]
\centerline{\epsfig{file=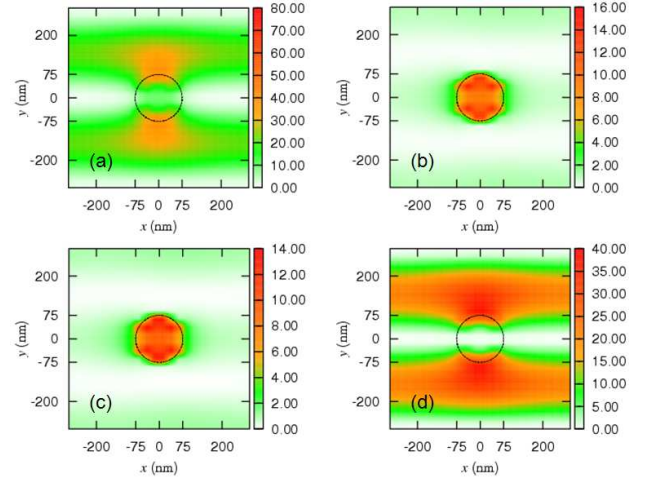,width=0.95\linewidth}}
\caption{(Color online) Electric field density on $z = 0$ of the $s$-excited
($0\pm 1$) SPP mode, $k_x = 0.1275$ and $\omega = 0.9849$.  The symmetry along
$x=0$ and $y=0$ indicates the localized nature of the resonance.}
\label{cylindrical:e:s:r75}
\end{figure}

In this section, we present the calculated field densities of the excited ($0,\pm
1$)-SPPs and discuss their symmetry properties.  A general discussion of the
symmetric properties of the eigenstates \citep{joannopoulos_1995,sakoda_2001}
is given in Appendix \ref{symmetric.operation}.
Electric field density ($|\vec{E}(\vec{r},\omega)|^{2}$) under $p$- and
$s$-polarized excitations is shown in Figs.  \ref{cylindrical:e:p:r75}(a)--(d)
and Figs. \ref{cylindrical:e:s:r75}(a)--(d), respectively.  The colors
indicate field strength, which ranges from 0--80 arbitrary units (A.U.).  The
cavity is outlined with black dotted lines.
As can been seen in Fig. \ref{cylindrical:e:p:r75}(a), strong electric fields are
distributed along the $x$-axis ($y=0$), with the strongest being
concentrated near the rim of the cavity.  This is in accordance with
Eqs. (\ref{even.odd.explicit.ex})--(\ref{even.odd.explicit.ez}), where
$E^2_{x}$ [Fig. \ref{cylindrical:e:p:r75}(b)] and $E^2_{z}$ [Fig.
\ref{cylindrical:e:p:r75}(d)] are symmetric along the $x$-$z$ plane, and
$E^2_{y}$ [Fig. \ref{cylindrical:e:p:r75}(c)] is anti-symmetric.  Therefore,
$E^2_{y}=0$ along $y=0$ and $y=\pm a/2$.  In contrast, the field is
asymmetric along the $x=0$ plane, which is due to the propagating nature of the
excited SPP mode \citep{iu_2008}.  When calculating the
$|\vec{E}(\vec{r},\omega)|^{2}$, two kinds of electric fields are present in
the system: one is the electric field of the resonant mode, labeled
$\vec{E}_{\rm res}(\vec{r},\omega)$, and the other is the electric field of
the incident light, labeled $\vec{E}_{\rm inc}(\vec{r},\omega)$.  According
to group theory, these two electric fields should have the same symmetric
properties under $C_{1h}$ symmetric operations.
The $p$-polarized light contains an electric field at the incident plane (the
$x$-$z$ plane) whose time average is always symmetric along the $y=0$ plane,
but asymmetric along the $x=0$ plane.  As a result, the resultant field
density, $|\vec{E}(\vec{r},\omega)|^{2}$, is asymmetric along the $x=0$ plane.

To further investigate the $p$-excited $(0,\pm 1)$ SPP, we show in
Fig. \ref{cylindrical:exy:field:p:r75} a snapshot of the field:
$E_x(\vec{r},t)$ and $E_y(\vec{r},t)$ during the FDTD simulations.  The
symmetry conditions, $E_x(x,y,z)=E_x(x,-y,z)$
[Eq. (\ref{even.odd.explicit.ex})] and $E_y(x,y,z)=-E_y(x,-y,z)$
[Eq. (\ref{even.odd.explicit.ey})], can be observed clearly in the figures,
although $E_y$ is too week and can barely be observed. It should be noted
that the $x$-component of the electric field is very weak at $x=0$, and is mainly
concentrated at ($x=\pm r$), which suggests that the local field should, at least
to the first order, be dominated by point dipole-like distribution, (that is,
$E_x\propto p_x)$, where $p_x$ is the total dipole moment along the $x$-axis.
This is in accordance with the hypothesis of \citet{abajo.prb.2005}, who treats
each cavity as a resonator that consists of a simple capacitor, an inductor and
resistance.

\begin{figure}[htbp]
\centerline{\epsfig{file=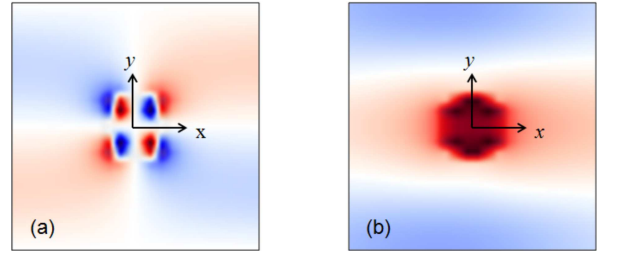,width=0.9\linewidth}}
\caption{(Color online) Snapshots of the electric field distribution of the
$s$-excited SPP mode in Fig. \ref{cylindrical:e:s:r75}.}
\label{cylindrical:exy:field:s:r75}
\end{figure}
We next turn our attention to the $s$-excited ($0,\pm 1$) SPP modes,
$|\vec{E}(\vec{r},\omega)|^{2}$, under $s$-polarized excitation, as shown in
Figs. \ref{cylindrical:e:s:r75}(a)--(d).  According to
Eqs. (\ref{even.odd.explicit.ex})--(\ref{even.odd.explicit.ez}), $E_x$ and $E_z$
should be anti-symmetric and $E_y$ should be symmetric along $y=0$, which
can be observed in Figs. \ref{cylindrical:e:s:r75}(b)--(d).  It should be
noted that the field densities of $E^2_x$ and $E^2_y$ are very similar.
However, as is revealed in Figs. \ref{cylindrical:exy:field:s:r75}(a)--(b), $E_x$
and $E_y$ are actually very different from each other.  Moreover, it should
also be noted that $E_y$ can be attributed to an effective dipole along the
$y$-axis.  In other words, $E_y\propto p_y$, which is similar to the case for
$E_x$ in the $p$-excited ($0,\pm 1$) mode [Figs. \ref{cylindrical:e:p:r75}(a) and
\ref{cylindrical:exy:field:p:r75}(a)].

\subsection{Conclusion}\label{discussion}
The present analysis relies only on the knowledge of $C_{4v}$-point group
symmetry, which should be valid for any plasmonic structure that belongs to the
same point group.  Therefore, our argument can be applied to all
$C_{4v}$-point group structures, such as square cavities, cross-shape grooves,
circular crafters, pyramidal structures
\citep{chuang.opt.express.2008,chen.opt.express.2009}, and
bottle-shape cavities \citep{li_apl_2008,iu_2008}.  Additionally, SPP
excitation along the $\Gamma$-M direction also has similar symmetric
properties \citep{sakoda_2001}, and the corresponding eigenmodes can also be
classified into even and odd modes, which are excited by $p$- and
$s$-polarized light accordingly.

To sum up, we have performed FDTD analysis on a gold cylindrical cavity array
and investigated the excited ($0,\pm 1$) mode based on the symmetry argument.  Although
very close in resonance frequencies,  the even and odd modes display very
different types of behavior in terms of field distribution, which has important
implications for near-field applications such as SERS.  It is hoped that this
work will provide insight into the fields of plasmonics and nano-optics.

\appendix
\section{Symmetric operations of eigenmodes}\label{symmetric.operation}
The eigenfunctions, $\vec{H}_{\vec{k}}(\vec{r},\omega)$, of a  periodic
plasmonic structure at eigenfrequency $\omega$ are Bloch functions that
satisfy the master equation \citep{joannopoulos_1995,sakoda_2001}:
\begin{eqnarray}
\hat{\Theta}\vec{H}_{\vec{k}}(\vec{r},\omega)
&=&
\nabla\times
\left (
\frac{1}{\varepsilon(\vec{r})}\nabla\times
\right )
\vec{H}_{\vec{k}}(\vec{r},\omega)\nonumber\\
&=& 
\frac{\omega^2}{c^2}\vec{H}(\vec{r},\omega),
\label{master}
\end{eqnarray}
where $\vec{k}$ denotes wavevectors within the first Brillouin zone.  Consider
a mirror transformation operator, $\hat{M}_{xz}$, along the $x$-$z$ plane that is mapped
from $\mathbb{R}^{3}$ to $\mathbb{R}^{3}$ by
\begin{eqnarray*}
\tilde{x} &=& \hat{M}_{xz}x = x,\\
\tilde{y} &=& \hat{M}_{xz}y = -y,\\
\tilde{z} &=& \hat{M}_{xz}z = z.
\end{eqnarray*}
Under this transformation, the $y$-coordinate obtains a minus sign, and the $x$-
and $z$-coordinates remain the same.  The vector fields are transformed by
operator $\hat{O}_{xz}$, defined as
\begin{equation}
\hat{O}_{xz}(f(\vec{r})\vec{F}(\vec{r}))=f(\hat{M}_{xz}\vec{r})\hat{M}_{xz}(\vec{F}(\hat{M}_{xz}\vec{r})),
\end{equation}
and thus,
\begin{eqnarray}
\hat{O}_{xz}f(\vec{r})   &=&  f(x,-y,z),    \label{fcn_trans}\\
\hat{O}_{xz}E_{x}(x,y,z) &=&  E_{x}(x,-y,z),\label{e_trans_x}\\
\hat{O}_{xz}E_{y}(x,y,z) &=& -E_{y}(x,-y,z),\label{e_trans_y}\\
\hat{O}_{xz}E_{z}(x,y,z) &=&  E_{z}(x,-y,z).\label{e_trans_z}
\end{eqnarray}
However, vectors defined by cross products are transformed
differently. For example, for magnetic field $\vec{B}$,
\begin{eqnarray*}
\vec{F} &=& q\vec{v}\times\vec{B},  \\
F_{x}\hat{x}+F_{y}\hat{y}+F_{z}\hat{z} &=& \left|
\begin{array}{ccc}
 \hat{x} & \hat{y} & \hat{z}       \\
 v_{x} & v_{y} & v_{z} \\
 B_{x} & B_{y} & B_{z} \\
\end{array}
\right| \\
&=& (v_{y}B_{z}-v_{z}B_{y})\hat{x}\\
&+& (v_{z}B_{x}-v_{x}B_{z})\hat{y}\\
&+& (v_{x}B_{y}-v_{y}B_{x})\hat{z}.
\end{eqnarray*}
To preserve the transformation properties of the force
and velocity, the magnetic field is transformed as follows
\begin{eqnarray}
\hat{O}_{xz}B_{x}(x,y,z) &=& -B_{x}(x,-y,z),\label{b_trans_x}\\
\hat{O}_{xz}B_{y}(x,y,z) &=&  B_{y}(x,-y,z),\label{b_trans_y}\\
\hat{O}_{xz}B_{z}(x,y,z) &=& -B_{z}(x,-y,z).\label{b_trans_z}
\end{eqnarray}
Electric fields and magnetic fields are transformed in different
ways under the mirror transformation.
\ \\

It is well known that $\hat{\Theta}$ commutes with $\hat{O}_{xz}$  for
periodic structures:
\begin{eqnarray*}
\hat{\Theta}\hat{O}_{xz} = \hat{O}_{xz}\hat{\Theta}.
\end{eqnarray*}
Also, if $\vec{k}=k\hat{x}$, then we have
\begin{equation*}
\hat{O}_{xz}\vec{k} = \vec{k},
\end{equation*}
and
\begin{equation}
\hat{O}_{xz}\vec{H}_{\vec{k}}(x,y,z) =
\pm\vec{H}_{\vec{k}}(x,y,z).
\label{even.odd}
\end{equation}
That is, the eigenmode is either symmetric ($+$) or antisymmetric ($-$) under
mirror transformation along the $x$-$z$ plane.  The substitution of
Eqs. (\ref{e_trans_x})--(\ref{e_trans_z}) and
(\ref{b_trans_x})--(\ref{b_trans_z}) into Eq. (\ref{even.odd}) gives
\begin{eqnarray}
E_{\vec{k},x}(x,y,z) &=& \pm E_{\vec{k},x}(x,-y,z),\label{even.odd.explicit.ex}\\
E_{\vec{k},y}(x,y,z) &=& \mp E_{\vec{k},y}(x,-y,z),\label{even.odd.explicit.ey}\\
E_{\vec{k},z}(x,y,z) &=& \pm E_{\vec{k},z}(x,-y,z),\label{even.odd.explicit.ez}\\
H_{\vec{k},x}(x,y,z) &=& \mp H_{\vec{k},x}(x,-y,z),\label{even.odd.explicit.hx}\\
H_{\vec{k},y}(x,y,z) &=& \pm H_{\vec{k},y}(x,-y,z),\label{even.odd.explicit.hy}\\
H_{\vec{k},z}(x,y,z) &=& \mp H_{\vec{k},z}(x,-y,z) \label{even.odd.explicit.hz}
\end{eqnarray}
for eigenmodes that satisfy Eq. (\ref{even.odd}).
In addition, for $p$-polarized lights,
$\vec{E}=(E_x,0,E_z)$, $\vec{H}=(0,H_y,0)$, and
\begin{eqnarray}
E_x(x,y,z) &=& E_x(x,-y,z),\nonumber\\
E_z(x,y,z) &=& E_z(x,-y,z),\nonumber\\
H_y(x,y,z) &=& H_y(x,-y,z).\label{p.explicit}
\end{eqnarray}
In contrast,
$\vec{E}=(0,E_y,0)$ and $\vec{H}=(H_x,0,H_z)$
for $s$-polarized lights; therefore
\begin{eqnarray}
E_y(x,y,z) &=& E_y(x,-y,z),\nonumber\\
H_x(x,y,z) &=& H_x(x,-y,z),\nonumber\\
H_z(x,y,z) &=& H_z(x,-y,z).\label{s.explicit}
\end{eqnarray}
A comparison of Eqs. (\ref{p.explicit}) and (\ref{s.explicit}) to
Eqs. (\ref{even.odd.explicit.ex})--(\ref{even.odd.explicit.hz}) suggests the $p$-
and $s$-polarized lights are symmetric and antisymmetric to the $x$-$z$ plane
mirror transformation, respectively.  Therefore, the symmetric (+) modes can
be excited only by $p$-polarized light that contains a symmetric electric field
along the $x$-$z$ plane.  Similarly, antisymmetric modes ($-$) can be
excited only by $s$-polarized light.

\section*{ACKNOWLEDGMENTS}
The authors acknowledge the assistance of S. H. Lee, Stephen Chan, and Frank Ng, and
thank Jensen Li and Z. H. Hang for their useful discussions of this work.
J.T.K.W. also acknowledges the support of S. S. Lam and T. L. Wan.
The finite-difference time-domain simulations were performed using the MIT-MEEP
package ver. 0.10, and computation was performed using CUHK's
high-performance computing (HPC) facility.  This work is supported by
the Research Grants Council of the Hong Kong Special Administrative Region,
China (Project no. CUHK/402807, CUHK/402908, and CUHK/403308).


\end{document}